%% file: main.tex
\let\iffinal\iffalse
\documentclass{sigplanconf}

\usepackage[utf8]{inputenc}
\usepackage[T1]{fontenc}

\def\lnkcolor{blue}
\iffinal \def\lnkcolor{black} \fi

\usepackage{hyperref}
\hypersetup{bookmarks=true,
            bookmarksopen=true,
            bookmarksdepth=2,
            colorlinks=true,
            pdftitle={%
Formal Proofs of Transcendence for e and pi as
an Application of Multivariate and Symmetric Polynomials},
            linkcolor=\lnkcolor,
            citecolor=\lnkcolor,
            urlcolor=\lnkcolor}

\usepackage{titlesec}
\titlespacing\section{0pt}{8pt plus 2pt minus 2pt}{3pt plus 2pt minus 2pt}
\usepackage{microtype}

\usepackage{url}
\usepackage{amsfonts,amsmath,mathtools}
\usepackage{todonotes}
\usepackage{alltt}
\usepackage{paralist}
\usepackage{algpseudocode}
\usepackage{algorithm}

\def\mathclap#1{\text{\hbox to 0pt{\hss$\mathsurround=0pt#1$\hss}}}

\newtheorem{lemma}{Lemma}
\newcommand{\bk}{\char'134}

\input{prelude}

\begin{document}

\toappear{}

\title{Formal Proofs of Transcendence for $e$ and  $\pi$ as
an Application of Multivariate and Symmetric Polynomials 
\thanks{This work was partly supported by the FastRelax
(ANR-14-CE25-0018-01) project of the French National Agency
for Research (ANR), and by a grant from the Cofund
Action AMAROUT II (\#291803).}}

\authorinfo{Sophie Bernard, Yves Bertot,\\{Laurence Rideau}}
{Inria}
{\{Sophie.Bernard,Yves.Bertot,Laurence.Rideau\}@inria.fr}
\authorinfo{Pierre-Yves Strub}
 {IMDEA Software Institute}{pierre-yves@strub.nu}

\maketitle 
\pagestyle{empty}
\thispagestyle{empty}

\begin{abstract}
We describe the formalisation in Coq  of a proof  that 
the numbers $e$ and  \(\pi\) are transcendental.
This proof lies at
the interface of two domains of mathematics that are often considered
separately: calculus (real and elementary complex analysis) and
algebra.  For the work on calculus, we rely on the Coquelicot library
and for the work on algebra, we rely on the Mathematical Components
library.  Moreover, some of the elements of our formalized proof originate
in the more ancient library for real numbers included in the Coq
distribution.  The case of \(\pi\) relies extensively on properties
of multivariate polynomials and this experiment was also an occasion
to put to test a newly developed library for these multivariate
polynomials.
\end{abstract}


\keywords Coq, Proof Assistant, Formal Mathematics, Transcendence,
Multivariate Polynomials

\section{Introduction}

As the ratio between areas of circles and squares, the number \(\pi\)
is known since antiquity, and the question of finding a method to 
construct this formula using purely algebraic tools, like a straight
edge and a compass has interested many mathematicians in history.
However, this algebraic approach did not make any progress, so that other
techniques, based on successive approximations, had to be summoned to
achieve better knowledge of this number.  So in the end, knowledge
on \(\pi\) is mostly a result of calculus rather than a result of algebra.

The history of the number \(e\) goes less far back in history,
although some of the properties of the number and the exponential
function can be inferred from a careful study of results known for
conics, more precisely hyperbolas, which were already studied
in antiquity, e.g. by Apollonius.  According to \cite{Coolidge50}, a
good historical account of this number should mention Apollonius (3rd
century B.C.), G. de St. Vincent (1647), Wallis, Mercator, Newton,
Leibniz. It is Euler that gave its name to the number, and in
particular, Euler defined \(e\) as follows:
\[e = 1 + \frac{1}{1} + \frac{1}{1\cdot 2} + \frac{1}{1\cdot 2\cdot 3}
+ \cdots\]

The concept of transcendental number appeared at the end of the
seventeenth century in the work of Leibniz.  Liouville was first to
prove the existence of transcendental numbers in 1844 and to produce a
constant that he proved transcendental.  The first case where
a known number was proved transcendental comes with Hermite's work
in 1874 \cite{Hermite74}, who proved that \(e\) is transcendental.
Lindemann then proved \(\pi\) to be transcendental.
This solves the question of constructing \(\pi\) with a straight edge
and compass, by simply stating that such a construction is impossible.
In this paper, we construct a formally verified proof of transcendence
for both numbers, relying on a proof produced by Niven \cite{Niven39}.

As a first experiment,
we formalized a proof of irrationality of \(\pi\), using a simple
proof also described by Niven \cite{Niven47}.  This work on the
irrationality of $\pi$ won't be described here, but it
helped us understand and develop the tools
needed for such a proof.  In particular, this initial experiment 
gave incentives for additions in the Coquelicot
library \cite{BolLelMel15}, which were later instrumental.

These three proofs (irrationality of \(\pi\) and transcendence of $e$
and \(\pi\)) use the same methodology \cite{Niven39}.  Using the
hypothesis that the number is rational or algebraic, we deduce an equality
$E_p = E'_p$, where both
\(E_p\) and \(E'_p\) are expressions depending on some integer \(p\).
We then show that, for \(p\)
sufficiently large, \(|E_p|\) must be
smaller than \((p-1)!\) for analytic arguments, and then that
\(|E'_p|\) must be an integer larger than \((p-1)!\) by using
algebraic and arithmetic arguments.

For the proof of transcendence of \(\pi\) it is
quite difficult to show that \(E'_p\) is an integer.  In this case, we
start with a polynomial \(B_\pi\) with integer coefficients such that
\(B_\pi(i\pi)=0\).  We manage to prove that \(E'_p\) is a symmetric
multivariate polynomial with integer coefficients applied to the roots
of \(B_\pi\). Using a general result about
multivariate polynomials known as the {\em fundamental theorem of
symmetric polynomials} we obtain that \(E'_p\) is obtained by
applying another polynomial with integer coefficients to the
coefficients of \(B_\pi\).

The previous paragraph shows that multivariate and symmetric
polynomials play a crucial role in this proof of transcendence.  Our
description of multivariate polynomials is built on top of the
Mathematical Components library \cite{GG4C,GGaFT}.  This library
already provides many of the notions of algebra that are needed for
our purposes.  It starts with basic mathematical structures such as
groups, rings, and fields, it provides univariate polynomials and
notions of algebraic numbers.  Most of these notions were needed for
the formalized proof of Feit and Thompson's odd-order
theorem \cite{GGaFT}.  However, multivariate polynomials were not
described in this context and our work is filling this gap.

In this paper, we first give an overview of the proofs, expressed in
mathematical terms.  We then illustrate the techniques used to make
the proofs verifiable by the Coq proof assistant \cite{Coq}.

{The formal development is available at the following address:
\url{http://marelledocsgit.gforge.inria.fr}.
The multinomial library is available as a separate component on
github\footnote{\url{https://github.com/strub/multinomials-ssr}}.

\section {Mathematical context}
In this section, we first give more mathematical background about
multivariate polynomials and we then concentrate on the proofs of
transcendence, showing that there is central lemma that can be
specialized for \(e\) and \(\pi\).

\subsection{Multivariate and symmetric polynomials}
\label{subsec:mpolymath}

Multivariate polynomials are polynomial expressions where several
indeterminates appear.  Common mathematical notations rely on writing
the indeterminates as \(X\),  \(Y\), and \(Z\) when there are no more
than three indeterminates and as \(X_i\) where \(i\) ranges between
\(1\) and \(n\) in the general case.  
The set of multivariate polynomials for a given set of indeterminates
and coefficients in a given ring also has the
structure of a ring.
We define {\em monomials} as expressions of the form
\(X_1^{k_1}\cdots X_n^{k_n}\) where \(k_i \in \mathbb{N}\) is the {\em degree in variable
\(X_i\)} and the sum \(k_1 + \cdots +k_n\) is called the {\em total
  degree}.  For a polynomial, the total degree is the maximum total
degree of its monomials.
We can sort the monomials of a multivariate polynomial
lexicographically, so that the first monomial is the one with the
highest degree in \(X_1\), and among the monomials of highest degree
in \(X_1\), the one with the highest degree in \(X_2\), etc.  In what
follows, we will call this first monomial in lexicographical order the
{\em leading monomial}.

The polynomials that are unchanged when permuting
the variables are called {\em symmetric} polynomials.  For
instance, the following polynomial is symmetric.
\[X^3Y + X^3 Z + XY ^3 + XZ ^ 3 + Y^3Z + Z^3Y\]

Among the symmetric polynomials, 
the polynomials with only the
permutations of a single monomial, with degree at most
one in each variable are called {\em elementary
  symmetric polynomials}.  For \(n\)-variable polynomials, there are
exactly \(n+1\) elementary symmetric polynomials, one for each total
degree between \(0\) and \(n\).
In this paper, we shall write \(\sigma_{n,k}\) for the elementary
symmetric polynomial of \(n\) variables with degree \(k\).
For three variables, the non-trivial elementary symmetric polynomials
are:
\[\sigma_{3,1} = X + Y + Z \quad 
\sigma_{3,2}= X Y + Y Z + X Z \quad \sigma_{3,3} =X Y Z.\]
The elementary symmetric polynomials are especially interesting in
proofs of transcendence because of Vieta's formula.  When working in
an algebraically closed field, considering
a polynomial \(P\) whose roots are \(\alpha_i\) (\(i \in \{1\ldots
n\}\)), noting \(c\) the leading coefficient, and noting \(\alpha\)
the vector whose components are
the roots \(\alpha_i\), this polynomial can be written as 
\[P = c \prod_{i=1}^n(X - \alpha_i) = c (\sum_{i=0}^n (-1)^ {n - i}
\sigma_{n,n-i}(\alpha) X ^ i)\] Thus, the elementary symmetric polynomials
applied to the roots of \(P\) give the coefficients of \(P\), after
multiplication by the leading coefficient and a sign factor.

Elementary symmetric polynomials make it possible to generate all
symmetric polynomials using multiplication, addition and
multiplication by a coefficient in the underlying ring.  This result
is known as the {\em fundamental theorem of symmetric polynomials}.
This theorem is instrumental in our formal proof.

Before stating and giving a precise proof to this theorem, we need to
introduce a last important notion, the notion of weight.  For a given
monomial \(M\), the {\em weight} \(w(M)\) of this monomial is the
weighted sum of the degrees in each variable, weighted by the rank of
this variable.
\[ w(X_1^{k_1}\cdots X_n^{k^n}) = \sum_{i=1}^{n} i\times k_i.\]
Intuitively, the weight of a monomial makes it possible to compute the
total degree of the result when applying this monomial to the sequence
of symmetric polynomials.

For instance, \(w(X_1^2X_2X_3) = 1\times 2 + 2 \times 1 + 3 \times 1 = 7\) and
applying the monomial
\(X_1^2X_2X_3\) to the symmetric polynomials leads to this result:
\[\sigma_{3,1}^2\sigma_{3,2}\sigma_{3,3}=(X + Y + Z)^2\times (XY + XZ
+ YZ)\times XYZ\]
This polynomial indeed has total degree 7.
\begin{lemma}\label{fundamental_theorem}
For every symmetric polynomial \(P\) in \(n\) variables
there exists a polynomial \(T\) in \(n\) variables
so that 
\[P[X_1;\cdots;X_n] = T[\sigma_{n.1}[X_1;\cdots;X_n];\cdots;\sigma_{n,n}[X_1;\cdots;X_n]].\]
Moreover, the weight of polynomial \(T\) is smaller than or equal to
the total degree of polynomial \(P\).
\end{lemma}

\noindent{\sl Proof.}  This proof is performed by well-founded
induction following the lexicographic order of the leading
coefficient.

Because \(P\) is symmetric, its leading monomial
necessarily has the shape \(X_1^{k_1}\cdots X_n^{k_n}\) with
\(k_1 \geq k_2 \cdots \geq k_n\).  This monomial naturally has
a total degree which is smaller than or equal to
the total degree of \(P\).  This monomial
is also the leading monomial of the expression:
\[\sigma_{n,1}^{k_1-k_2}\sigma_{n,2}^{k_2-k_3}\cdots
\sigma_{n,n}^{k_n}\]
Let us note temporarily \(M = X_1^{k_1-k_2}X_2^{k_2-k_3}\cdots
X_n^{k_n}\) (\(M\) is an \(n\)-variable polynomial),
then the above expression can also be written as follows:
\[ M[\sigma_{n,1};\cdots;\sigma_{n,n}]\]
if \(c\) is the coefficient of the leading monomial in \(P\), we can
consider the polynomial \[Q=P - c\times
M[\sigma_{n,1};\cdots;\sigma_{n,n}]\]
If \(Q\) is constant, then the required polynomial is \(T=cM-Q\) and
the weight of \(P\) is the weight of \(cM\).  If 
\(Q\) is non constant, it is symmetric (because it is the subtraction
of a symmetric polynomial from a symmetric polynomial) and its leading
coefficient is lexicographically less than the leading coefficient of
\(P\).  We
can deduce by induction hypothesis that there exists a polynomial
\(T'\) such that
\[Q[X_1;\cdots;X_n]=T'[\sigma_{n,1}[X_1;\cdots;X_n];\cdots;\sigma_{n,n}[X_1;\cdots;X_n]].\]
The weight of \(T'\) is less than or equal to the total degree of
\(Q\), which is less than or equal to the total degree of \(P\).
  The required polynomial is 
\[T = T' + c \times M\]
The weight of this polynomial is less than or equal to the maximum
of the weights of \(T'\) and \(M\), which is enough to guarantee the
weight property. $\square$

\subsection {Common lemmas for proofs of transcendence}

The two transcendence proofs are quite similar, so to factorize the
formalization work, we have proved a generalized lemma, that can then
be applied for both cases ($e$ and $\pi$).

The proof revolves around an analytic part and an algebraic part.  The
analytic part relies on the following integral:
\[I_{P}(\alpha) = \int_0^1 \alpha e^{-\alpha x} P(\alpha x){\sf d}x\]

We also consider an operation on polynomials, which adds up all the derivatives of this polynomial:
\[P_d = \sum_{i=0}^{deg P} P^{(i)},\]
  where $P^{(i)}$ denotes the i-th derivative of $P$ and \(deg P\) is the degree of \(P\).
\subsubsection{Preliminary results on integrals}

\begin{lemma}\label{I_identity}
For every polynomial \(P\) and every complex number \(\alpha\),
the following identity holds:
\begin{equation}
I_P(\alpha) = P_d(0)-e^{-\alpha}P_d(\alpha) \label{eq1}
\end{equation}
\end{lemma}
\noindent{\sl Proof.}  Note that $P^{(0)} = P$, $P^{(i)} =  0 $ for
 $deg P < i $, and \(e^{\alpha\times 0}=1\).
\begin{eqnarray*}
\lefteqn{(P_d(\alpha x)e^{-\alpha x})' =}\\
&=  & -\alpha e^{-\alpha x}P_d(\alpha x) + \alpha e^{-\alpha x} P_d'(\alpha x)\\
& = & \alpha e^{-\alpha x}( P_d'(\alpha x) - P_d(\alpha x))\\
& = & \alpha e^{-\alpha x}\left(\sum_{i=1}^{deg P + 1} P^{(i)}(\alpha x) - \sum_{i=0}^{deg P} P^{(i)}(\alpha x)\right)\\
& = & \alpha e^{-\alpha x} (P^{(deg P+1)}(\alpha x) -  P^{(0)}(\alpha x))\\
 & = & - \alpha e^{-\alpha x}P(\alpha x)
\end{eqnarray*}
Hence we have:
\[
\mathclap{\int_ 0^1}\; \alpha e^{-\alpha x}P(\alpha x){\rm d}x = [-P_d(\alpha x)e^{-\alpha x}]_0^1= P_d(0)-e^{-\alpha}P_d(\alpha)~\square \]
\subsubsection{Main construction}
In what follows, we work with a sequence of \(n\) non-zero complex numbers 
\(\alpha_i \in \mathbb{C}^*\) and a non-zero integer \(c\).  We first consider
the following polynomial:
\begin{equation}
T = c (X-\alpha_0)(X-\alpha_1)\cdots (X-\alpha_{n-1})  \label{hyp2}
\end {equation}
We then introduce, for an arbitrary \(p\), (which we shall later choose to be very large and prime), the polynomial \(F_p\) :

\[ F_p = X^{p-1}T^p. \]

And as in formula (\ref{eq1}), we consider \(I_{F_p}\)  and the sum of derivatives of \(F_p\), 
which we shall note \(F_{pd}\).  We also consider an intermediate lemma concerning solely the derivatives of order larger than or equal to \(p\).

\begin{lemma}\label{derive_n_p_multiple_fact_p}
When the coefficients of \(T\) are integers, the coefficients of 
\[\sum_{i = p}^{deg F_p} F_p^{(i)}\]
are divisible by \(p!\)
\end{lemma}

\noindent{\sl Proof.}
For every polynomial \(Q\) with integral coefficients, all
coefficients of \(Q^{(m)}\) are divisible by \(m!\)  So in
particular, all coefficients of \(F_p^{(i)}\) are divisible by \(i!\)
and if \(p \leq i\), they are divisible by \(p!\) Divisibility can then
be inherited by the sum \(\square\)

We introduce a family of polynomials \(G_p\)
defined as follows:
\[G_p = \frac{\sum_{i=p}^{deg F_p} F_p^{(i)}}{p!}\]
Thanks to Lemma~\ref{derive_n_p_multiple_fact_p}, the coefficients of \(G_p\) will inherit from \(T\) the property that
the coefficients are integers.

\subsubsection{Main lemma}\label{main-lemma-sec}
We can now state our main lemma, which relies on the concepts of \(T\), \(F_p\), and \(G_p\), which we just introduced.

\begin{lemma}\label{main_lemma}
If \(T\) has integral coefficients, then for every sequence \(\gamma_i\) of integers and any \(k\in \mathbb{Z}^*\), if the following equality holds
\[
 k + \sum_{i=0}^{n-1}\gamma_i e^{\alpha_i} = 0
\]
then for \(p\) prime and large enough,
\[c^{np}\sum_{i=0}^{n-1} \gamma_i G_p(\alpha_i) \not\in {\mathbb{Z}}\]
\end{lemma}

\noindent{\sl Proof.}
Let us assume that \(T\) has integral coefficients and that 
\begin{eqnarray}
 k + \sum_{i=0}^{n-1}(\gamma_i e^{\alpha_i}) = 0 \label{hyp1}
\end{eqnarray}

Consider the sum \(\sum -\gamma_i e^{\alpha_i}I_{F_p}(\alpha_i)\) and re-use the result from (\ref{eq1}).
\begin{eqnarray*}
\sum_{i=0}^{n-1} -\gamma_i e^{\alpha_i}I_{F_p}(\alpha_i)&=&
\sum_{i = 0}^{n-1}- \gamma_i e^{\alpha_i}(F_{pd}(0) - e^{-\alpha_i}F_{pd}(\alpha_i))\\
& = & \sum_{i = 0}^{n-1}- \gamma_i e^{\alpha_i}F_{pd}(0)+ \sum_{i = 0}^{n-1}\gamma_i F_{pd}(\alpha_i)
\end {eqnarray*}
by (\ref{hyp1}) we have $\sum- \gamma_i e^{\alpha_i} = k$, and therefore we get the following result:
\begin{equation}
\sum_{i=0}^{n-1} -\gamma_i e^{\alpha_i}I_{F_p}(\alpha_i)=
kF_{pd}(0) +  \sum_{i = 0}^{n-1}\gamma_iF_{pd}(\alpha_i)\label{Eqc}
\end{equation}

We will now consider a new equation, where both sides are multiplied by
\(c^{np}\).
\begin{equation}
c^{np}\sum_{i=0}^{n-1} -\gamma_i e^{\alpha_i}I_{F_p}(\alpha_i)=
c^{np} kF_{pd}(0) +  c^{np}\sum_{i = 0}^{n-1}\gamma_iF_{pd}(\alpha_i)\label{Eqc_cnp}
\end{equation}
We will call \(E_p\) the left hand side and \(E'_p\) the right hand side.
We will now show that, for $p$ large enough, the  \(|E_p|\) is less than
$(p - 1)!$.

Let $A$ be an upper bound of the finite sequence $|\alpha_i|$.
The following inequalities hold:
\begin{eqnarray}
|I_{Fp}(\alpha_i)| &=&
 \left|\int_0^1 \alpha_i e^{-\alpha_i x} F_p(\alpha_i x){\sf
 d}x\right|\nonumber \\
& \leq &\int_0^1\left| \alpha_i e^{-\alpha_i x} F_p(\alpha_i
 x)\right|{\sf d}x
\nonumber \\
& \leq & A e^A\int_ 0^1|F_p(\alpha_ix)|{\sf d}x \nonumber\\
&  \leq & A e^AA^{p-1}\int_ 0^1|T(\alpha_ix)|^p{\sf d}x \label{maj_IFp}
\end{eqnarray} 
by definition of $F_p$.

Let $M$ be an upper bound of $|T(\alpha_ix)|$ on $[0,1]$.
Then we get $|I_{F_p}(\alpha_i)| \leq A^pe^AM^p$ and there exist
\(K\) and \(L\) (independent of \(p\)) such that $|E_p = c^{np}\sum- \gamma_i e^{\alpha_i}I_{F_p}(\alpha_i)|\leq K L^{p-1}$.

We know that $\lim_{n \rightarrow \infty} x^n/n! = 0$, so when \(p\) is large enough  $|E_p| < (p-1)!$

From now on, we want to show that for \(p\) prime and large
enough, \(|E'_p|\) is larger than or equal to \((p-1)!\)

The numbers \(\alpha_i\) are roots of \(F_p\) with multiplicity \(p\), and therefore \(F^{(m)}(\alpha_i) = 0\) for every \(m < p\).  This gives a way to reduce the sum of all derivatives of \(F_p\) in the numbers \(\alpha_i\).
\begin{eqnarray*}
{F_{pd}(\alpha_i)}
 &=& {\sum_{j=0}^{p-1} F_p^{(j)}(\alpha_i) + \sum_{j=p}^{deg F_p} F_p^{(j)}(\alpha_i)}\\
&=& 0 + {\sum_{j=p}^{deg F_p} F_p^{(j)}(\alpha_i)}
\end{eqnarray*}
Similarly for the value of \(F_{pd}\) in 0, we know that 0 is a root of multiplicity \(p-1\) of \(F_p\), and thus \(F_p^{(m)}(0)=0\) for every \(m < p-1\).  We can thus write the following equations:
\begin{eqnarray*}
{F_{pd}(0)}
 &=& {\sum_{i=0}^{p-2} F_p^{(i)}(0) + F_p^{(p-1)}(0) + \sum_{i=p}^{deg F_p} F_p^{(i)}(0)}\\
&=& 0 +  (p-1)!T(0)^p + {\sum_{i=p}^{deg F_p} F_p^{(i)}(0)}
\end{eqnarray*}
Putting these facts together, we can reason on \(E'_p\).
\begin{eqnarray*}
E'_p&=&c^{np}\sum_{i=0}^{n-1} -\gamma_i e^{\alpha_i}I_{F_p}(\alpha_i)\\
&=&c^{np} k F_{pd}(0) +  c^{np}\sum_{i = 0}^{n-1}\gamma_iF_{pd}(\alpha_i)\\
&=&
c^{np}\times k\times(p-1)!\times T(0)^p + p!G_p(0))\\
&&{\strut{}  + p!c^{np}\sum_{i=0}^n\gamma_i G_p(\alpha_i)}
\end{eqnarray*}

When \(c^{np}\sum\gamma_i G_p(\alpha_i)\) is an integer, the right hand side
is obviously divisible by \((p-1)!\).  But when \(p\) is a large enough prime
(larger than \(c\), \(k\), and \(T(0)\)) only the last two terms can be
divided by \(p!\)  This expression is an integer divisible
by \((p-1)!\) but not by \(p!\).  the absolute value must be larger than or equal to \((p-1)!\)~\(\square{}\)

\subsection{Case of \(e\)}\label{math4e}

If \(e\) is an algebraic number, let \(A_e\) be a polynomial with
integer coefficients such that \(A_e(e)=0\).  Let us note \(a_i\) the
coefficients of \(A_e\)
\[A_e = \sum_{i=0}^{n_e} a_i X \qquad A_e(e) = \sum_{i=0}^{n_e}a_i e^i\]
Without loss of generality, we can assume that 0 is not a root
of \(A_e\) (i.e. \(a_0 \neq 0\)).  We instantiate the main lemma in
the following fashion:
\begin{enumerate}
\item \(\alpha_0 = 1, \alpha_1 = 2 \cdots \alpha_{n_e - 1} = n_e\).  The alphas are
non-zero complex numbers,
\item \(\gamma_0 = a_1 \cdots \gamma_{n_e - 1} = a_{n_e}\).  The
numbers \(\gamma_i\) are integers; \(k = a_0\) is a non-zero integer.
\item the polynomial \(T=(x-1)\cdots (x - n_e)\) trivially has integral
coefficients, we take \(c=1\).
\item The expression \(c^{np} \sum \gamma_i G_p (\alpha_i)\) trivially
is an integer because the numbers \(c\), \(\gamma_i\),
and \(\alpha_i\) are integers and the polynomial \(G_p\) has integer
coefficients.
\end{enumerate}

\subsection {Case of $\pi$}\label{math4pi}
For \(\pi\), we work with \(i\pi\) instead of \(\pi\) because it is
equivalent to prove that one is algebraic or the other.  Let \(B_\pi\)
be a polynomial of degree \(n_\pi\) with rational coefficients so that
\(B_\pi(i\pi)=0\).  Without loss of generality, we can assume that this polynomial has the form
\[B_\pi = \prod_{i=0}^{n_\pi - 1} (X -\beta_i).\]
where none of the \(\beta_i\) is zero and \(i\pi = \beta_0\).

To build the sequence \(\alpha\), we first build a sequence \(\alpha'\) and then remove the zero elements.
The sequence \(\alpha'\) is built from the sequence \(\beta\) by
taking all the sums of non-empty subsets of the sequence \(\beta\).
Thus \(\alpha'\) contains \(\beta_0\), \(\beta_1\), \(\cdots\),
\(\beta_0 + \beta_1\), \(\cdots\)
We should note that \(i\pi\) appears in the sequences
\(\alpha'\) and \(\alpha\).

We then rely on the Euler equation: \(e^{i\pi}=-1\).  Because \(i\pi\) appears in the sequence \(\beta\) we have the following property:
\[\prod (1 + e^{\beta_i}) = 0\]
But if we expand this product, we obtain a formula:
\[1 + \sum e^{\alpha'_j} = 0\]
Let \(n\) be the length of the sequence \(\alpha\) obtained by
removing the zero elements from the sequence \(\alpha'\) and
let \(k\) be one plus the number of elements in \(\alpha'\) that are zero.
\[k + \sum_{i = 0}^{n-1} e^{\alpha_i}=0\]
Taking \(\gamma_i = 1\) for every \(i\) we have:
\[k + \sum_{i = 0}^{n-1} \gamma_i e^{\alpha_i} = 0.\]
We can now instantiate the main lemma in the following fashion:
\begin{enumerate}
\item \(\alpha\) is a sequence of non-zero complex numbers,
\item \(\gamma_0 = \cdots = \gamma_n = 1\), the \(\gamma_i\) are
integers, \(k\) is a positive integer.
\item The polynomial
  \[T' = (X-\alpha_0)(X-\alpha_1)\cdots(X-\alpha_{n-1})\]
is such that
\[(X-\alpha'_0)(X-\alpha'_1)\cdots(X-\alpha'_{2^{n_\pi}-2}) = X ^{k-1}T' \]
By Vieta's formula, the coefficients of \(X^{k-1}T'\) are
obtained by applying elementary symmetric polynomials on \(\alpha'\),
but the components of \(\alpha'\) are stable modulo permutation of the
variables \(\beta_i\) and so these coefficients are also obtained by
applying symmetric polynomials on \(\beta\).
Using the fundamental theorem of symmetric polynomials for a first
time, these coefficients are obtained by applying multivariate
polynomials  with integer coefficients to the elementary symmetric polynomials
of \(\beta\), which
are the coefficients of \(B_\pi\).  Thus, the coefficients of \(T'\) are all
rational and we can exhibit a constant \(c\) so that
\[T = c T' = c  (X-\alpha_0)(X-\alpha_1)\cdots(X-\alpha_{n-1})\]
has integral coefficients.
\item  Because \(\gamma_i = 1\) for every \(i\), the following formula is
actually symmetric in the \(\alpha_i\)
\[c^{np}\sum_{i=0}^{n-1} \gamma_i G_p(\alpha_i)\]
Since we know that \(G_p\) is a polynomial with integral
coefficients, and using the fundamental theorem of symmetric
polynomials a second time, we can deduce that this expression can be
computed by applying a polynomial with integer coefficients on the coefficients
of \(c\prod (X - \alpha_i)\), which also are integers.
\end{enumerate}

\section {Formalization}
Our description of the formal development is organized into five parts:
we describe the working context provided by the Mathematical
Components library, the development of multivariate polynomials,
the connection to the Coquelicot library, the
proof of the main lemma, and the specialization to \(\pi\).

\subsection {The common infra-structure: Mathematical Components}\label{mathcomp}
The mathematical Components library draws from a tradition that was
started with the formal proof of the four-color theorem in 2004
\cite{GG4C} and continued for the proof of the Feit-Thompson theorem
\cite{GGaFT}.

This library is built on top of the Coq proof assistant, an interactive
theorem proving tool that relies on type theory: most theorem
statements are types in a functional programming language and
theorems are functions from one type to the other.  One of the key
functionalities of Coq that we use intensively is the mechanism of
{\em canonical structures}.

When compared to other libraries for the Coq system, the Mathematical
Components library also differs in its use of {\em small-scale
  reflection},
 where most decidable logical expressions are represented by
 boolean formulas.  When compared with libraries developed for other interactive
proof systems, we should note that classical logic-based interactive
theorem provers, like Isabelle \cite{Nipkow-Paulson-Wenzel:2002} or
HOL-Light
\cite{HOL-Light} do represent
logical statements with
boolean formulas, but on the other hand, they have practically no tool
to distinguish between a constructive proof and a non-constructive
one.

For mathematical practice, the Mathematical Components library
provides an extensive collection of notations, which are compatible
with the level of abstraction.  For instance {\tt \(a\) + \(b\)} represents
the addition of two elements {\em in any ring}.  So that we will write
the same formula {\tt \(a\) + \(b\)}, whether \(a\) and \(b\)
represent one-variable or multivariate polynomials, even though these
are two different types.
In the same vein, the Mathematical Components library also provides
support for big iterated operations such as
\[\sum_{i=0}^{n-1}\gamma_ie^{\alpha_i}\]
which represent the repeated iteration of the addition operator of any
ring structure.  In our formal development, this formula is written
as:
\begin{coq}
\sum_(0 <= i < n) (gamma i)
\end{coq}
The tag \verb"%:~R" is needed here to express that the numbers
\(\gamma_i\) which are integers should be injected into the type of complex 
numbers.   This is a first illustration that 
in terms of notations, not everything is seamless.  Another illustration is
that we need to
accommodate different notions of exponentiation.  In rings,
exponentiation is only defined when the exponent is a natural number,
by iterating the multiplication operation, in plain fields (like
\(\mathbb{Q}\)) it is also
defined for negative integral exponents, by simply taking
multiplicative
inverses,
but in complete fields like \(\mathbb{R}\) or \(\mathbb{C}\), it is
defined as the result of completely different processes, by taking the
limit of a power series (for which a convergence proof is actually provided).
For this reason, we use a different notation
for each of these four cases and it is sometimes tricky to reconcile two
formulas that use different notations.

The Mathematical Components library is designed to
alleviate the difficulties arising from the distinction between
various kinds of numbers.  The first step is to describe the various
levels of properties that a set and operations on this set can
satisfy.  At the first level, the notion of monoid (only one binary
law, which is associative and has a neutral element), then groups,
then commutative groups, then rings, integral domains, fields, up to
algebraically closed and complete fields.  Each structure at a higher
level in the ladder satisfies the properties of the lower rungs (and
hence the theorems can be written once and for all).  The Mathematical
Components library then ensures that the abstract properties can be
recognized and used for a wide class of concrete instances.  Thus,
integers can be recognized as a ring structure.  Then many theorems
are expressed in terms of morphisms.  Actually, the ring of integers is even
recognized as the initial ring, so that the notation \verb"%:~R"
is defined as a morphism from the
type of integers to any ring (for instance to the ring of complex
numbers, or to the ring of polynomials with integer coefficients).

The Mathematical Components library defines the notion of algebraic
number, by considering two fields and a morphism from the first to the
second field.  A number \(a\) in \(F\) is
algebraic over \(E\) and \(\phi\), if there is a polynomial with
coefficients in \(E\), \(\sum_i p_i X^i\) such that the equality
\(\sum_i \phi(p_i) a^i = 0\) holds.  So the notion of algebraic number
is not intrinsic, it relies on two fields and a morphism.  In our
experiment, the common meaning of algebraic numbers correspond to
numbers that are algebraic over \(\mathbb{Q}\) in \(\mathbb{C}\),
using the natural morphism of rational numbers into the field of
complex numbers as the morphism.  So we write {\tt algebraicOver ratr x},
where {\tt ratr} is the morphism from rational numbers to any field.

Our work on multivariate polynomials integrates in this context.
These polynomials are described by choosing a data-structure and then
providing the operations of addition, opponent, multiplication, that
make this type a ring, together with
the operations that are expected from a polynomial: decomposing into
the monomial basis (and thus observing the coefficients), evaluating
at a multiplet of values, finding the leading monomial, etc.  We show
that many operations actually are morphisms.  For instance, the
function for evaluating at a given multiplet is a ring morphism from
the ring of multivariate polynomials to the ring of coefficients.

\subsection{Formalizing Multivariate Polynomials}

In this section we describe the construction of multivariate
polynomials. In the univariate case, \ssr concretely represents
polynomials as a sequence of coefficients. For instance, the
univariate polynomial $\sum_{i \in \mathbb{N}} \alpha_i X^i$, where
$\{ \alpha_i \}_i$ is a null set, is represented by the sequence
$[\alpha_0, \ldots, \alpha_k]$, with $\alpha_k \ne 0$ being the last
non-null coefficient when ordering the $\alpha_i$s by their indices.

This representation can be generalized to a finite number of
indeterminates, using an enumeration
$\{ (\epsilon^1_i,\ldots,\epsilon^n_i) \}_{i \in \mathbb{N}}$ of the
countable set $\mathbb{N}^n$. In this case, the multivariate
polynomial

\begin{center}
  \[\sum_{i_1,\ldots,i_n} \alpha_{i_1,\ldots,i_n} X_1^{i_1} \cdots X_n^{i_n}
  = \sum_{i \in \mathbb{N}} \underbrace{\alpha_{\epsilon^1_i,\ldots,\epsilon^n_i}}_{\beta_i}
      X_1^{\epsilon_i^1} \cdots X_n^{\epsilon_i^n}\]
\end{center}

\noindent would be represented by the sequence $[\beta_0, \ldots, \beta_k]$,
where $\beta_k$ is again the last non null coefficient when taken
in the ordering implied by the enumeration of $\mathbb{N}^n$.
However, although this representation is quite canonical and effective
in many respects in the univariate case (e.g. the degree and the
coefficients of a polynomial are easily definable from its sequence of
coefficients), it seems less effective in the multivariate case. For
instance, the link between the $\beta_i$s and
$\alpha_{i_1,\ldots,i_n}$s is now non trivial and depends on the
enumeration of $\mathbb{N}^n$, which may lead to overly complicated
definitions. Moreover, this representation does not lift canonically
to the case of an infinite number of indeterminates, and would require
an ad-hoc construction, for example by taking a structure limit.

Other options are available to represent multivariate polynomials. For
example, one could iterate the univariate case, representing
$k[X_1 \cdots X_n]$ as $k[X_1]\cdots[X_n]$. Yet again, this
representation does not lift canonically to an infinite number of
variables, and equipping this representation with the canonical
structures of the \ssr algebra hierarchy requires some
contortions. Indeed, one can not simply define the type of
multivariate polynomials as:

\begin{coq}
Fixpoint mpoly (R : ringType) (n : nat) :=
  if n is p.+1 then {poly (mpoly R p)} else R.
\end{coq}

\noindent as \ls${poly (mpoly R p)}$ requires \ls$(mpoly R p)$
to be already equip\-ped with a \ls$ringType$ structure at the
time of definition. A proper definition is:

\begin{coq}
Fixpoint mpoly (R : ringType) (n : nat) : ringType :=
  if n is p.+1 then
    [ringType of {poly (mpoly R p)}]
  else R.
\end{coq}

However, this construction triggers a complexity explosion in the \Coq
unification algorithm, and we never succeeded in equipping
\ls$(mpoly (R : idomainType) p)$ with an integral domain structure.

Last, the \ssr trunk archive comes with an embryo theory for
multivariate polynomials~\cite{GGaFT} that relies on the quotient of a
free algebra (built from the indeterminates, the ring constants and
the uninterpreted operators \ls$+$ and \ls$*$) by an interpretation
relation, using univariate polynomials as the domain of
interpretation. This definition directly handles the case of an
infinite number of indeterminates and is appealing as it allows the
definition of basic functions via the manipulation of ring expressions
---~once these manipulations are proved to be compatible with the
quotient relation. However, the quotient relation is non-trivial and
we expect these proofs of compatibility to be harder that necessary.

\medskip

For our formalization, we take the angle of representing a
multivariate polynomial literally as a monoid ring, i.e. as a formal
sum of the form $\sum \alpha_i (X_1^{k^1_i} \cdots X_n^{k^n_i})$.
For that purpose, we develop an independent structure for free abelian
groups. We then obtain a structure for $k[X_1,\ldots,X_n]$ by
instantiating the one of free abelian groups, taking the coefficients
from the multivariate polynomials base ring $k$ and using a free
commutative monoid over $\{1,\ldots,n\}$ as the monoid of
generators. This gives us the basic structure for multivariate
polynomials in $n$ variables. We then develop an extensive library,
including polynomials derivation, evaluation, morphisms, and prove
that our representation is isomorphic (as a ring) to the iterated
representation. We then use our development to prove the fundamental
lemma of symmetric polynomials, proving that the ring of symmetric
polynomials in $n$ variables is isomorphic to the ring of polynomials
in the elementary symmetric polynomials.

\subsubsection{Free Abelian Groups}

The core structure of our multivariate polynomials library is the one
of free abelian groups. The latter is based on the \ls$quotient$
libraries of \ssr.
Assume $T$ to be a set of generators and $G$ to be a group. The free
abelian group with coefficients in $G$ and generators in $T$ is the
set of formal sums of the form $\sum_{x \in T} \alpha_x\ (x)$, where
$\{\alpha_x\}_{x \in T} \in G^T$ is a null set, equipped with a group
structure where $0 = \sum_x 0\ (x)$ and
$\sum_x \alpha_x\ (x) + \sum_x \beta_x\ (x) = \sum_x (\alpha_x +
\beta_x)\ (x)$.

\smallskip

The formal sum $\sum_{x \in T} \alpha_x\ (x)$ can be represented by
the finite map $\{ x \mapsto \alpha_x \mathrel{|} \alpha_x \ne 0 \}$,
and, in term of formalization, by the association list
$[(x, \alpha_x) \mathrel{|} x \in T, \alpha_x \ne 0]$. We define
\ls$prefreeg$ as the type of all valid association lists, i.e. as the
collection of all sequences \ls$s$ of type \ls$seq (T * G)$ s.t. no
pair of the form \ls$(_, 0)$ occurs in \ls$s$ and for any
\ls$(x : T)$, a pair of the form \ls$(x, _)$ appears at most once in
\ls$s$.

\begin{coq}
Definition reduced (g : seq (T * G)) :=
     (uniq [seq zx.1 | zx <- g])
  && (all  [pred zx | zx.2 != 0] g).

Record prefreeg : Type := mkPrefreeg {
  seq_of_prefreeg : seq (T * G);
  _ : reduced seq_of_prefreeg }.
\end{coq}

The intent of \ls$prefreeg$ is to give a unique representation of a
free-group expression, up to the order of the coefficients. For
instance, if $g = k_1\ (x_1) + \cdots + k_n\ (x_n)$ (with all the
$x_i$s pairwise distinct and all the $k_i$s in $G$), then the
\ls$reduced$ sequence \ls$s = [:: (x_1, k_1), ..., (x_n, k_n)]$, or
any sequence equal to \ls$s$ up to a permutation, is a valid
representation of $g$. The type \ls$freeg$ of free abelian groups is
then obtained by taking the quotient of \ls$prefreeg$ by the
\ls$perm_eq$ equivalence relation. Such a construction requires the
axiom of choice over \ls$seq (T * G)$, which amounts to both \ls$T$
and \ls$G$ equipped with the \ls$choiceType$ structure, as provided in
the Mathematical Components.

\medskip

We now show that our representation is faithful. Given a
sequence \ls$(s : seq (T * G))$ $= [(k_i, x_i)]_i$ (not necessarily
reduced), we define \ls$precoeff x s$ as:

\begin{coq}
Definition precoeff x s : G :=
  \sum_(k <- s | k.1 == x) k.2.
\end{coq}

The function \ls$precoeff$ computes the coefficient of $x$ in the
formal sum $\sum_i k_i\ (x_i)$. (Note that if the sequence \ls$s$ is
reduced, this amounts to looking up in \ls$s$, defaulting to \ls$0$ is
\ls$x$ cannot be found). We then prove that two \ls$freeg$ inhabitants
are equal if and only if they agree on their coefficients:

\begin{coq}
Lemma freegP (g1 g2 : freeg T G) : reflect
  (forall x, precoeff x g1 = precoeff x g2) (g1 == g2).
\end{coq}

On the other hand, we show that we can, from any association sequence,
not necessarily reduced, construct a reduced one s.t.  \ls$precoeff$
agrees on both.

\begin{coq}
Definition reduce (s : seq (T * G)) := ...
Lemma rdce_reduce s : reduced (reduce s).
Lemma rdce_eq s x : precoeff x s = precoeff x (reduce s).
\end{coq}

From there, we equip the type \ls$freeg$ with a group structure, the
representative of $0$ being the empty list, and the representative of the
sum being the reduced concatenation of the representatives of the
operands. We also define all the usual notions related to free groups
(domain, coefficient, degree, ...), most of them being defined using
the following group morphism:

\begin{coq}
Definition fglift
  (M : lmodType R) (f : T -> M) (g : freeg T G) : M :=
    \sum_(k <- repr g) k.1 *: (f k.2).
Definition deg (g : freeg T int) : int :=
  fglift (fun x => 1) D.
\end{coq} 

\noindent where \ls$fglift$ stands for the group homomorphism
defined by:

\begin{center}
  \ls$fglift$ $(f, \sum_{x} \alpha_x\ (x)) = \sum_{x} \alpha_x\ f(x)$
\end{center}

\subsubsection{Multivariate Polynomials}

We construct our type of multivariate polynomials in $n$ variables by
instantiating the type of free abelian groups, taking for the
coefficients the base ring and using a free commutative monoid over
$\{1,\ldots,n\}$ as the set of generators.

We use the type of finite functions from $\{1,\ldots,n\}$ to
$\mathbb{N}$ for representing the free commutative monoid over
$\{1,\ldots,n\}$. This type is then equipped with the adequate monoid
structure.

\begin{coq}
Inductive multinom := Multinom of {ffun 'I_n -> nat}.
Coercion fun_of_multinom m := let: Multinom m := m in m.
Definition mzero := Multinom [ffun _ => 0].
Definition madd m m' := Multinom [ffun i => m i + m' i].
\end{coq}

The type \ls$mpoly R n$ of multivariate polynomials in $n$ variables
over the ring $R$ is then defined as \ls$freeg R (multinom n)$. It
directly inherits the basic functions (domain, coefficients, ...) and
structures (decidable equality, group) of \ls$freeg$. It remains to
prove that our structure is a ring, defining the ring laws as follows:

\begin{coq}
Definition pone := << 1 *g 0
Definition pmul (p q : mpoly R n) :=
  \sum_(k1 <- dom p) \sum_(k2 <- dom q)
    << (coeff p k1) * (coeff p k2) *g (k1 + k2)
\end{coq}

\noindent where \ls$<< k *g x >>$ is the free group notation
for $k\ (x)$ and \ls$0
operations.
Looking at these definitions, we see that our definitions follow
closely the ones that can be found in textbooks. The subtle difference
relies in the ranges of the summations. Most textbooks make implicit
that these sums are finite, whereas in our case, this is made explicit
by iterating over the domains of \ls$p$ and \ls$q$. Besides this
explicit management of summation supports, the proofs follow closely
the ones of textbooks. Moreover, we tackle the difficulty of manual
handling of summations supports by adapting the \emph{big
  enough}\cite{cohen_phd} mechanism to our case, allowing us to defer
and to compute \emph{a posteriori} the domain on which the involved
summations must be done.

From there, we develop an extensive library, including polynomial
differentiation, evaluation, morphisms. The development follows closely the
one of the univariate case, and we equip our constructions with the
relevant algebraic structures that can be found in the \ssr library.
In the next section, we illustrate our library by describing the formal
proof of the fundamental lemma of symmetric polynomials.

\subsubsection{Formalizing the Fundamental Lemma}

The \emph{fundamental lemma of symmetric polynomials} states, for any
symmetric polynomial $p$ in $n$ variables, the existence and
uniqueness of a \emph{decomposition} polynomial $t$ s.t. $p$ is equal
to $t \circ \{ X_i \mapsto \sigma_{n,i} \}$ where the $\sigma_{n,i}$s
are the $n$ elementary symmetric polynomials defined by
$\sigma_{n, k} =
   \sum_{s \in \mathcal{P}(\{1,\ldots,n\})}^{\#|s| = k}
     \left( \prod_{i \in s} X_i \right)$.{}\footnote{The notation
  $p \circ \{ X_i \mapsto q_i \}$ is the multivariate polynomial
  composition, returning the polynomial obtained by the formal
  substitution of the $X_i$s by the $q_i$s in $p$.}
We detail in this section the \emph{existence} part of the proof of
the fundamental lemma, as stated by the following \Coq statement:

\begin{coq}
Definition symmetric := [forall s, msym s p == p].

Definition mesym (k : nat) : {mpoly R[n]} :=
  \sum_(h : {set 'I_n} | #|h| == k) \prod_(i in h) 'X_i.

Let S := [tuple (mesym i) | i < n].

Lemma sym_fundamental (p : {mpoly R[n]}) :d
  p \is symmetric -> { t | t \mPo S }.
\end{coq}

\noindent where \ls$msym s p$ stand for $p$ where the indeterminates
are permuted by the permutation \ls$s$, and \ls$p \mPo S$ stands for
the composition of $p$ with the $n$ elementary symmetric polynomials
\ls$(mesym i)$.  All these definitions follow closely the mathematical
ones and rely essentially on facilities already present in the \ssr
library and in our multivariate polynomials library. Due to the lack
of space, we do not expand on these definitions and we immediately
shift to the proof formalization.

\smallskip

Following Section~\ref{subsec:mpolymath}, the proof of the fundamental
lemma constructively builds the decomposition polynomial, using some
fixed well-founded monomial ordering for enforcing termination.
In our formalization, we explicitly exhibit this function, and prove
that its terminates and is correct w.r.t. the fundamental lemma
statement.
For that, we define\footnote{The notation $\ominus$ stands for the
  truncated subtraction, returning $0$ if the result of the
  subtraction is negative.} in Algorithm~\ref{fig:symf} a
\emph{procedure} \textsc{symf}, that given a multivariate polynomial
$p$, computes, if it exists, the decomposition $t$ of $p$ s.t.
$p = t \circ \{ X_i \mapsto \sigma_{n, i} \}$. As written, this
procedure may not terminate, and this is indeed the case when $p$ is
not symmetrical.

\begin{algorithm}
\begin{algorithmic}[1]
\Function{symf1}{\mbox{$p$ : polynomial in $n$ variables}}
\If {$p$ is the zero polynomial} \Return $(0, 0)$
\Else
\State $\alpha\ (X_1^{k_1} \cdots X_n^{k_n})
  \gets \mbox{ the leading monomial of } p$
\State $m \gets X_1^{k_1 \ominus k_2} \cdots X_{n-1}^{k_{n-1} \ominus k_n} X_n^{k_n}$
\State \Return $(\alpha\ m, p - \alpha\ (m \circ \{ X_i \mapsto \sigma_{n, i} \}))$
\EndIf
\EndFunction
\State
\Function{symf}{\mbox{$p$ : polynomial in $n$ variables}}
\State $(t, p) \gets (0, p)$
\While {$p$ is not the zero polynomial}
\State $(q, p) \gets $ \Call{symf1}{$p$}
\State $t \gets t + q$
\EndWhile
\State \Return $t$
\EndFunction
\end{algorithmic}
\caption{\label{fig:symf}}
\end{algorithm}

Our main result in this section is the formalization of the
termination and correctness of the \textsc{symf} procedure.
We start by translating the pseudo-code of the decomposition to some
\Coq function. Unsurprisingly, the unbounded \textbf{while} loop
cannot be defined in Coq as-is.  Instead, we code a variant \ls$symfn$
of \textsc{symf} that takes an extra (fuel) parameter $n$ and returns
the pair $(t, p)$ that results after the execution of exactly $n+1$
iterations of \textsc{symf}:

\begin{coq}
Definition symf1 (p : {mpoly R[n]}) :=
  if p == 0 then (0, 0) else
    let (a, A) := (coeff p (mlead p), mlead p) in
    let m := [multinom i < n | (A i - A i.+1)
    (a *: 'X_[m], p - a *: ('X_[m] \mPo S)).

Fixpoint symfn (k : nat) (p : {mpoly R[n]}) :=
  if k is k'.+1 then
    let (t1, p) := symf1 p in
    let (t2, p) := symfn k' p in (t1 + t2, p)
  else symf1 p.
\end{coq}

Beside the \ls$mlead p$ construct, that returns the maximum (leading)
monomial of $p$ for a fixed monomial ordering, \ls$symf1$ uses only
standard functions of the multivariate polynomials library. For the
definition of \ls$mlead$, we develop a library for orders. Notably, we
define the lexicographic lift of orders to fixed size tuples, and
proved that totality and well-foundness are preserved. This allows use
to define some monomial ordering as:

\begin{coq}
Definition mnmc_le (m1 m2 : 'X_{1..n}) :=
  lex [posetType of nat] (mdeg m1 :: m1) (mdeg m2 :: m2).
\end{coq}

\noindent where \ls$lex$ is the lexicographic lift of an order (here
the one of natural numbers) and \ls$m1$, \ls$m2$ are cast, via a
coercion, to their respective $n$-tuple indeterminate powers.
This order differs from the one defined in
Section~\ref{subsec:mpolymath} by prepending the degree of the
monomials to the sequence of indeterminates powers. It is known as the
\emph{degrevlex} monomial order and guarantees that a monomial is
always strictly larger than any monomial of strictly lower degree.
Being the lexicographic lift of a total well-founded order, we obtain
that \ls$mnmc_le$ is also a total well-founded order over monomials,
and equip this with the relevant structure of the order library.
Notably, this allows us to define a new induction principle for
polynomials, based on the monomial ordering:

\begin{coq}
Lemma mleadrect (P : {mpoly R[n]} -> Prop) :
  (forall p,
     (forall q, (mlead q < mlead p)
  -> forall p, P p.
\end{coq}

We are now ready to define the \emph{leading coefficient} of a
polynomial $p$ as its greatest monomial (i.e. with a non-null
coefficient) for the monomial order we just defined:

\begin{coq}
Definition mlead p := (\max_(m <- msupp p) m)
\end{coq}

The \emph{leading coefficient} denomination here makes sense as
\ls$mlead p$ returns one of the monomials of \ls$p$ of maximal degree.

\paragraph{Correctness and completeness}
We then prove that \ls$symf1$ is correct w.r.t. the fundamental lemma,
and that this function progresses w.r.t. the monomial ordering. The
first property is needed to prove the correctness of the final
decomposition function, whereas the latter is used when proving the
termination of the iteration of \ls$symf1$:

\begin{coq}
Lemma symf1P (p : {mpoly R[n]}) : p \is symmetric ->
  [&& ((symf1 p).2 == 0) || (mlead (symf1 p).2 < mlead p)
    , (symf1 p).2 \is symmetric
    & p == (symf1 p).1 \mPo S + (symf1 p).2].
\end{coq}

In essence, the property states that, when given a symmetric
polynomial $p$, \ls$symf1$ returns a pair $(t, q)$ composed of a
partial decomposition polynomial $t$ and a remainder $q$
\emph{smaller} than $p$, i.e. such that
$p = t \circ \{ X_i \mapsto \sigma_{n,i} \} + q$, and s.t. $q$ is
symmetric and is either null or with a leading monomial strictly lower
than the one of $p$. Moreover, in case of a null remainder, we see
that the problem is solved, as we obtain a polynomial $t$ s.t.
$p = t \circ \{ X_i \mapsto \sigma_{n,i} \}$.

Having \ls$symf1$ returning a symmetric polynomial, this property can
be directly lifted to \ls$symfn$ by induction on the \emph{fuel}
argument. Again, as for \ls$symf1$, if the remainder returned by
\ls$symfn$ is null, then the problem of decomposing the input
polynomial is solved. We prove that \ls$symfn$ is complete, i.e. that
it returns a null remainder as long as it is given enough fuel, and
give a concrete bound for the needed fuel:

\begin{coq}
Lemma symfnS (p : {mpoly R[n]}) :
  { n : nat | p \is symmetric -> (symfn n p).2 = 0 }.
\end{coq}

The property is directly proved using the \ls$mleadrect$ induction
principle and by application of the progress property. At this point,
we have all the ingredients to define the final decomposition function
(by hiding the fuel argument of \ls$symfn$) and to give a formal proof
of the fundamental lemma:

\begin{coq}
Definition symf (p : {mpoly R[n]}) :=
  (symfn (tag (symfnS p)) p).1.

Lemma symfP (p : {mpoly R[n]}) :
  p \is symmetric -> p = (symf p) \mPo S.
\end{coq}

Making the decomposition function explicit  has one major benefit: the
decomposition algorithm is not hidden inside some proof term, and we can
prove, \emph{a posteriori}, extra properties on the decomposition
polynomial by a simple induction over the fuel argument of \ls$symfn$
---~the fact that the decomposition polynomial has a weight smaller
that the degree of the input polynomial is proved this way.

\subsection {Analysis part}
For the analysis part, we rely on the axiomatic real numbers provided
by the Coq distribution and the extension provided by the Coquelicot
library \cite{BolLelMel15}.  The main advantage of this library is
that it treats integration and differentiation in a smooth way.

\subsubsection{Bridges between various formalized complex numbers}
Complex numbers were developed independently in Coquelicot and the
Mathematical Components.  In the Coquelicot
library, complex numbers are used for
real analysis.  In the Mathematical Components,
the construction of complex numbers is studied from a more abstract
point of view as the result of field extensions on top of arbitrary
real closed fields.  In particular, the construction in Mathematical
Components makes it possible to consider the field of complex
algebraic numbers (which is not topologically complete but where equality is
decidable) independently
from the field of complex numbers (which is topologically complete, but where
equality is not decidable).

We need to re-use many of the notions from both
libraries: for instance, the
concept of algebraic number, inherited from Mathematical Components
and the concept of integrating a function from \({\mathbb{R}}\) to
\(\mathbb{C}\) inherited from Coquelicot.

In a first module, which we call {\tt Rstruct}, we instantiate most of
the structures of Mathematical Components on the type of real numbers,
whose description comes from the Coq distribution.
In our {\tt Rstruct} module, we show that as a
consequence of the set of axioms that describe the real numbers, this
type also satisfies many of the characteristics required for
Mathematical Components types.  For instance, the
Mathematical Components has a notion of {\tt eq\_type}, a type where
equality is mirrored by a boolean test function.  The existence of
this boolean test function is a consequence of the axioms defining the
real numbers.

It should be noted that Mathematical Components structures require
ring structures to satisfy a choice
property.  This choice property is not a consequence of the axioms
defining {\tt R} in the Coq distribution, so we do not only use these
axioms, but also the axiom provided in the {\tt Epsilon} package,
whose strength is similar to the axiom of choice.  For
a similar reason, we also rely on functional extensionality.

Thus, we are studying the consequences of the set of axioms that describe
the real numbers in Coq's standard library.  Even though Coq's logic is
constructive, our proof is classical because it relies on axioms that are only
valid in classical logic.

The Mathematical Components library provides a type
constructor named {\tt complex}, which takes as input any type equipped
with a {\em real closed field}
structure and returns a new type with a field structure, which is
algebraically closed and a field extension of the former.  We applied
this constructor on the type of real numbers and obtained a type 
we called {\tt complexR}.  This type
satisfies the properties of being a
{\tt numClosedFieldType}, in other words an algebraically closed
type with a norm satisfying triangular inequalities and stability with
respect to multiplication.

 We then needed to establish many
correspondences between this type of complex numbers and the similar
construction present in the Coquelicot library.  In the end, we manage
to combine the concepts of limits, derivability, continuity,
integrability from the Coq distribution and the Coquelicot library
with the algebraic properties of the complex numbers from the
Mathematical Components library.

\subsubsection{Subsets of complex-integers and complex-naturals}
For our proof, we also need to define the predicates {\tt Cnat} and
{\tt Cint} that recognize exactly the complex numbers which are
integers and natural numbers, respectively.
Once these
predicates are defined, the Mathematical library provides a natural
notion of polynomial over a predicate: it is polynomial such that all
coefficients satisfy the predicate.  For instance, to express that
polynomial {\tt P} has all its coefficients in {\tt Cint} one simply
writes:
\begin{coq}
P \is a polyOver Cint
\end{coq}

In the same manner that mathematical structures can be attached to
types, they can also be attached to predicates.  In the case of the
{\tt Cint} predicate, we show that 0 and 1 satisfy this predicate, and
that it is stable for all the ring operations.  This in turn makes
it possible to invoke general theorems provided once and for all
for all stable predicates.
For instance, our development contains the
following declarations about {\tt Cint}:
\begin{coq}
Fact Cint_subring : subring_closed Cint.
Proof. ... Qed.

Canonical Cint_opprPred := OpprPred Cint_subring.
Canonical Cint_addrPred := AddrPred Cint_subring.
...
Canonical Cint_subringPred :=
                SubringPred Cint_subring.
\end{coq}
For example, any statement of the form {\tt (x + y) \bk{}in Cint}
can be transformed into the two statements {\tt x \bk{}in Cint} and
{\tt x \bk{}in Cint} using a lemma {\tt rpredD} which was defined once
for all predicates compatible with addition.

\subsubsection{Formally defining complex exponential}
Complex exponential is defined from real exponential by using the
following definitions, when \(x\) and \(y\) are both real numbers.
\[e^{x + iy} = e^x \times (\cos y + i \sin y)\]

In our formal text, this definition is written in the following
manner:
\begin{coq}
Definition Cexp (z : complexR) :=
  (exp(Re z))
\end{coq}
In this formula, {\tt Re} and {\tt Im} denote the projections that
return the real and
imaginary part of a complex
number, \verb"%:C" denotes the injection from real numbers to complex
numbers, and the functions {\tt exp}, {\tt cos}, and {\tt sin} are
functions from real numbers to real numbers that were provided for a
long time in the Coq distribution (these last three functions are defined
analytically as limits of power series).

It is then fairly easy to prove all the relevant properties of
this exponential function, in particular the morphism property from
\((\mathbb{C}, +)\) to \((\mathbb{C^\star}, \times)\) and the Euler
equation \(e^{i\pi}=-1\).

For differentiation, we restrict our study to the differentiation of functions
from \(\mathbb{R}\) to \(\mathbb{C}\) and perform most of the study
using differentiation componentwise, thus viewing \(\mathbb{C}\) as a
\(\mathbb{R}\)-vector space of dimension 2.  We derive all the usual
properties of derivatives with respect to addition and multiplication.
For the differentiation of exponential, we only study
the function \(x \mapsto e^{ax}\) from \(\mathbb{R}\) to
\(\mathbb{C}\), where \(a\) is an arbitrary complex number.  We prove
the following two lemmas, which are enough for our proof.
\begin{coq}
Lemma ex_Crderive_Cexp (a : complexR) (x : R) :
  ex_derive (fun y : R_NormedModule =>
             Cexp(a * y
...
Lemma Crderive_Cexp (a : complexR) (x : R) :
  Crderive (fun y => Cexp(a * y
    a * Cexp(a * x
\end{coq}
The first of these two lemmas states that the function \(y\mapsto e
^{ay}\) is differentiable everywhere.  The second lemma gives the value of
the derivative.

For integration, we only define notions related to integration of
functions from \(\mathbb{R}\) to \(\mathbb{C}\), by considering
integration independently on each component.  
Unfortunately, we were unable to benefit from the Coquelicot library
when developing this part, because integration in vector space was
less smoothly designed than integration of real-valued functions.  For
instance, the Coquelicot library only provides theorems linking
integrals and antiderivatives for real-valued functions, and not for
more general functions with values in arbitrary normed spaces.  As an
illustration, we had to prove the theorem that links integration and
antiderivative:
\begin{coq}
Lemma RInt_Crderive f a b:
  (forall x, Rmin a b <= x <= Rmax a b ->
              ex_derive f x) ->
  (forall x, Rmin a b <= x <= Rmax a b -> 
         Crcontinuity_pt (Crderive f) x) ->
  CrInt (Crderive f) a b = f b - f a.
\end{coq}
In this statement, {\tt CrInt} represents the integral
operator for functions from \(\mathbb{R}\) to \(\mathbb{C}\), and
{\tt Crderive f} represents the derivative of {\tt f}.  A
first hypothesis expresses that {\tt f} must be differentiable everywhere
in the integration interval, a second hypothesis expresses that
the derivative must be continuous everywhere in this interval.

\subsubsection{Details on a proof at the frontier between libraries}
\label{small-proof}
A necessary elementary fact about exponentials is
that for any natural numbers \(a\) and \(b\), for \(n\) large enough,
\(a b^n < n!\).
This is a fact about exponentials because the power series \(\sum
x ^n /n!\) converges to \(e^x\) for any \(x\), and therefore
\(x^n/n!\) converges
towards 0.  We manage to perform this proof by re-using a general
theorem about the generic term of a converging series (provided by Coquelicot),
 and then the formal definition of exponential.

As an illustration of the formalization work, we can follow the steps
of the formal proof for the statement
\[\forall a, b\in\mathbb{N}, \exists M, M < n \Rightarrow a b^n < n!\]

We first state that we use notations for real numbers:
\begin{coq}
Open Scope R_scope.
\end{coq}
We then establish equalities relating notions of factorial on the one
hand, and notions of exponentiation on the other hand between the
Mathematical Components library and the Coq distribution.

We can start the proof of our result:
\begin{coq}
Lemma p_prop1 (a b : nat) : 
  exists M, forall n, (M <= n -> a * b ^ n < n`!)
\end{coq}
Here we use the qualifier {\tt ( \dots{} )\%N} to express that
the comparisons between {\tt M} and {\tt
N} and between \verb"a * b ^ n" must be read with notations for
natural numbers.

The first stage of our proof is to prove that the sequence \(a \times b
^n/n!\) has a limit of 0 in the real numbers.  This is written in the
following fashion:
\begin{coq}
have : is_lim_seq
  (fun n => INR a * (INR b ^ n / INR (fact n))) 0.
\end{coq}
In this statement, {\tt is\_lim\_seq} is a concept from Coquelicot
library that simply means that a function
from \(\mathbb{N}\) to \(\mathbb{R}\) has a given limit
in \(\overline{\mathbb{R}}\).  Also {\tt fact} is used to denote
the factorial function as defined in Coq's distribution, which is
different from the factorial function as defined in Mathematical
Components.  We also needed to add an equality lemma to reconcile the
two definitions.

The notation \(0\) refers to a constant of type \(\mathbb{R}\) but
{\tt is\_lim\_seq} expects a value of type \(\overline{\mathbb{R}}\)
(named {\tt Rbar} in the Coquelicot library).
The type reconstruction
algorithm of Coq detects this discrepancy and solves the problem by
introducing a coercion from \(\mathbb{R}\)
to \(\overline{\mathbb{R}}\).

This intermediate statement is proved in three lines of tactics;
the first two lines get rid of multiplication by \(a\), first showing
that \(0\) could also be viewed as a the multiplication \(a \times 0\)
in \(\overline{\mathbb{R}}\), and then applying a lemma about limits and
multiplication by a scalar constant ({\tt is\_lim\_seq\_scal}).
We then apply a lemma
stating that the general 
term of a converging series has limit 0 ({\tt ex\_series\_lim\_0}).
We then exhibit
the limit of \(\sum b^n/n!\), this is \(e^b\), written {\tt exp (INR b)} in Coq
syntax.  These are the two lines of development:
\begin{coq}
rewrite [_ 0](_ : _ = Rbar_mult (INR a) 0);
      last by rewrite /= Rmult_0_r.
apply/is_lim_seq_scal_l/ex_series_lim_0;
      exists (exp (INR b)).
\end{coq}
In the first line, the notation \verb"[_ 0]" is used to state that not
only the symbol 0 should be replace by \(a \times 0\), but also the
coercion that injects 0 into \(\overline{\mathbb{R}}\).

The third line then invokes a lemma
to connect power series from Coquelicot and power series from the Coq
distribution.  At this point, it happens that {\tt exp} is actually
defined as the first component of a dependent pair (a concept of type theory),
where the second component is the proof of convergence of the series.  
The theorem {\tt svalP} makes it possible to use this fact.
\begin{coq}
by apply/is_pseries_Reals; apply:svalP.
\end{coq}
The next two lines use the limit statement and specialize it to find
the value {\tt M'} so that \(|a b^n/n!-0| < 1\) for every \(n\) larger than 
{\tt M'}, express
that this is the required value for {\tt M}, and consider an arbitrary
{\tt m} that is larger than {\tt M'}.  This is written in the
following manner:
\begin{coq}
rewrite -is_lim_seq_spec => ils;
   case: (ils (mkposreal _ Rlt_0_1)) => M'.
rewrite /pos => PM'; exists M' => m /leP M'm;
   move: (PM' _ M'm).
\end{coq}
It takes 5 more lines to get rid of subtraction and the absolute
function, multiply both sides of the comparison
by \(m!\), reconcile the duplicate definition of factorial, and use
morphism properties to move the statement from
real numbers to natural numbers.  This proof still feels unnatural
because we spend too much time reconciling the various definitions.

\subsection {The common lemma}
The formal development for the common lemma is concentrated in one file
where we first study the notion of multiple root of order
\(m\) of a polynomial.
\begin{coq}
Definition mroot (p : {poly R}) m x :=
    rdvdp (('X - x
\end{coq}
This definition relies on the notion of polynomial
divisibility noted {\tt rdvp} provided by Mathematical Components.
Here, {\tt x\%:P} means the injection of the value {\tt x}
from the ring {\tt R} to the type of polynomials and ``\verb"^+"''
represents a power.

We show by induction on \(m\)
that if \(x\) is a root of
polynomial \(p\), it is a multiple root of order \(m\) of
polynomial \(p^m\).  We also show that if \(x\) is a multiple root of
order \(m\) of a polynomial \(p\), then it is also a root of
order \(m-i\) of the iterated derivative \(p^{(i)}\), for any \(i\)
smaller than \(m\).
\begin{coq}
Lemma mrootdP p m x :
  reflect (forall i : 'I_m, mroot p^`(i) (m - i) x)
     (mroot p m x).
\end{coq}
In this statement {\tt 'I\_m} denotes the type of natural numbers smaller
than {\tt m} and \verb"p^`(i)" denotes the
\(i^{th}\) derivative of polynomial {\tt p}.  Also, {\tt reflect} is a
specific form of equivalence statement used when the right-hand
formula is a boolean statement and the left-hand formula is a
proposition represented by a type.

We then start assuming the existence of a few
objects and a few hypotheses: the natural
numbers \(n\) and \(c\) (both non-zero), the sequence \(\gamma_i\)
of integers and the non-zero integer \(k\), and the
sequence \(\alpha_i\) of complex numbers, and we assume
that \(\alpha_i\) is non zero for every \(i\) such that \(0 \leq i <
n\).  We define \(T =c \times \prod_{i < n} (X - \alpha_i)\) and we
prove a variety of simple properties about it.  In particular,
we show that there is an upper bound
\(M_i\) for the value of the function \(x \mapsto T(\alpha_i x)\).
For instance, this is written in the following manner.

\begin{coq}
Lemma ex_Mc i :
 {M : R | forall x : R, 0 <= x <= 1 ->
                         norm T.[alpha i * x
...

Definition M i := sval (ex_Mc i).
\end{coq}
Elements of the type {\tt \{M : R | ...\}} are pairs of a value and a
proof that the value satisfies a given property, known as dependent pairs.
The function {\tt sval}
simply returns the first component of such a dependent pair.  
Lemma {\tt svalP} is used to return the second component of the pair and
express its statement in terms {\tt sval} (lemma {\tt svalP} was already
used in section~\ref{small-proof}).
We then use the numbers \(M_i\), \(|\alpha_i|\),
and \(|\gamma_i e^{\alpha_i}|\) to define well-chosen values \(a\)
and \(b\) so that \(a b^{(p-1)}\) will be a suitable upper bound of the 
sum of integrals we consider later in the proof.  We also combine
a few extra constraints on \(p\) which are expressed in the following
lemma:
\begin{coq}
Lemma p_prop2 :
  exists p : nat, prime p &&
      (a * b ^ p.-1 < (p.-1)`!) &&  (p > `|k|) &&
      (p >  `| floorC (T.[0])|) && (p > c).
\end{coq}
This time we don't use a dependent pair to express the existence of
{\tt p}.  The reason is that {\tt p\_prop2} relies on {\tt p\_prop1}
which was already expressed using an existential statement instead of
a dependent pair (because {\tt p\_prop1} was proved non-constructively).
Still, natural numbers can be enumerated and the property in
the existential statement is decidable, the Mathematical Components
library provides a function {\tt xchoose} that returns a suitable
witness.  This is how we define the natural number {\tt p} that will
play a central role:
\begin{coq}
Definition p := xchoose p_prop2.
\end{coq}
We then define \(F_p\), we state lemmas about its degrees and its
roots and we consider the operation of summing all the derivatives of
a polynomial, written as follows:
\begin{coq}
Definition Sd (P : {poly complexR}) j0 :=
  \sum_(j0 <= j < (size P)) P^`(j).
\end{coq}
In what follows, \(F_{pd}\) will be represented formally by {\tt Sd Fp
0}.

We do not prove lemma~\ref{I_identity}, but
only its specialization to the polynomial \(F_p\) and the
values \(\alpha_i\).  We actually define the value {\tt IFp i} to
represent \(I_{F_p}(\alpha_i)\) by taking directly the expression
\(F_{pd}(0) - e^{-\alpha_i}F_{pd}(\alpha_i)\)
\begin{coq}
Definition IFp i :=
  (Sd Fp 0).[0] - Cexp (-alpha i) * (Sd Fp 0).[alpha i].
\end{coq}
Then the computation of the integral is written as follows:
\begin{coq}
Lemma CrInt_Fp i :
  CrInt (fun x => alpha i * Cexp(-alpha i * x
                    * (Fp.[alpha i * x
\end{coq}
The proof of the bound~\ref{maj_IFp} is done step by step in a
sequence of lemmas that consider integrals with integrands that get
smaller and smaller.  These proof are quite tedious because for every
integral, we have to re-explain that this integral is well-defined.

We conclude this analysis part with the following upper bound on \(|E_p|\):
\begin{coq}
Lemma eq_ltp1 :
  `|(c ^ (n * p))
       \sum_(0 <= i < n)
             -(gamma i)
   < ((p.-1)`!)
\end{coq}
It should be noted that there was no need in this part to assume that
\(T\) or \(F_{p}\) have integer coefficients.

We then add the extra assumption that \(T\) has
integer coefficients.  We can derive \(F_p\)
also has integer coefficients.

In our mathematical exposition, we define \(G\) as the sum of the
derivatives of \(F_p\), starting from the derivative of order \(p\), and
divided by \(p!\).
In our development, we reuse a concept provided by Mathematical
Components, noted \verb"p^`N(n)" for a polynomial {\tt p} and
a natural number {\tt n} with the following property:
\begin{coq}
nderivn_def
     : forall (R : ringType) (n : nat) (p : {poly R}),
       p^`(n) = p^`N(n) *+ n`!
\end{coq}
In this lemma, the notation {\tt *+} is used to represent
multiplication by a natural number.
Another companion lemma expresses that
if {\tt P} has integer coefficients, then \verb"P^`N(k)" also does.  The
polynomial \(G\) is then defined as the sum of all derivatives of
\verb"Fp^`N(p)" up to order \(pn\) and \(G\) has
integer coefficients.

After adding the  assumption that \(c^{np}\sum \gamma_i
G(\alpha_i)\) is an integer, we prove the final steps of
Section~\ref{main-lemma-sec} by representing them as specific lemmas.
After discharging all assumptions, 
the main lemma actually has the following general statement:
\begin{coq}
main_contradiction :
forall n : nat, n != 0 -> forall k : int, k != 0 ->
forall (gamma : nat -> int) (c : nat), c != 0 ->
forall alpha : nat -> complexR, 
(forall i : 'I_n, alpha i != 0) ->
  \prod_(i < n) ('X - (alpha i)
            \is a polyOver Cint ->
(k
   \sum_(0 <= i < n)
        (gamma i)
~ (\sum_(0 <= i < n)
    (gamma i)
    c ^ (n * p n k gamma c alpha))
\end{coq}

\subsection {Instantiating the common lemma: cases of e and pi}
The proof of transcendence of \(e\) is a very simple instantiation of the
common lemma
and the formal development does not pose any difficult problem.  For
instance, the following line describes the \(\alpha\) sequence for \(e\):

\begin{center}
\ls$Definition alpha i := (i.+1
\end{center}

Still it takes a little work to let the Coq system accept that \(e^i=e^i\)
because \(e^i\) should be written {\tt Cexp i} in the formula
\(k + \sum \gamma_i e^i\), while it is written \(e^i\) in the
polynomial with integer coefficients that has \(e\) as root.

For \(\pi\), we first show that we can work with \(i\pi\), because the
{\tt polyOver} predicate is stable for the product operation. Then  we  obtain a
polynomial \(B_\pi\) with complex-integer coefficients that has \(i\pi\) as
root, a positive leading coefficient, and a non-zero constant coefficient.
Because {\tt complexR} has the structure
of an algebraically closed field, there is a function that returns a
list {\tt betaseq} containing all the roots of this polynomial.  This
is slightly different from Section~\ref{math4pi}, where we only expect
\(B_\pi\) to  have rational coefficients.  In what follows, we name {\tt s} the
degree of \(B_\pi\) minus one (it is also the size of {\tt betaseq}
minus one).

The statement that \(\prod_{i=0}^{s}(1 + e^\beta_i) = 0\) is written
as follows in our development:
\lstinline{\prod_(b <- betaseq) (1 + (Cexp b)) = 0}

In section~\ref{math4pi}, we describe the sequence \(\alpha'\)
directly as the sequence of all sums of non-empty subsets
of \(\beta\).  In the formal development, we follow a structure that
is closer to Niven \cite{Niven39} and we consider separately the sequences
of sums for any subset of \(\beta\) of cardinal \(j \leq n\).  These
sequences are named {\tt pre\_alpha' \(j\)}.  We then define the
sequences of non-zero elements from {\tt pre\_alpha' \(j\)}, these are
called {\tt pre\_alpha \(j\)}.  The sequence {\tt alpha} is finally
obtained by concatenating all the sequences {\tt pre\_alpha \(j\)}.

The next main step is to show that for each of the sequences {\tt
pre\_alpha' \(j\)}, there exists a polynomial with integer coefficients and
a non zero constant coefficient, that
has the elements of this sequence as roots:
\begin{coq}
Lemma alpha'_int (j : 'I_s.+1) :
  {c : nat |
   (\prod_(a <- pre_alpha' j)('X - a
       \is a polyOver Cint) && (c != 0

\end{coq}
This proof reduces to showing that applying elementary symmetric
polynomials on the sequences {\tt pre\_alpha' \(k\)} returns a symmetric
polynomial expression on {\tt betaseq}.  For this
we rely on a general lemma from the multivariate library about
composing symmetric polynomials with multiplets of polynomials that
are stable by permutation.
\begin{coq}
Lemma msym_comp_poly k
          (p : {mpoly R[n]})(t : n.-tuple {mpoly R[k]}):
     p \is symmetric
  -> (forall s : 'S_k,
        perm_eq t [tuple (msym s t`_i) | i < n])
  -> p \mPo t \is symmetric.
\end{coq}
In this statement, {\tt msym s \verb+t`_i+} represents the polynomial
\verb+t`_i+ where the variables have been permuted according to the
permutation {\tt s}.  Proving this premise in the case of {\tt
pre\_alpha'} is quite tedious.

Other key steps in the formalization are the uses of the {\em
fundamental theorem of symmetric polynomials}. This theorem applies
for every symmetric polynomial with coefficients in a commutative
ring, and returns a polynomial with coefficients in that ring.

In our case, the polynomial we are working on is a polynomial with
complex coefficients which satisfy the {\tt Cint} predicate.  If
we use directly the fundamental theorem on this polynomial, we
obtain a resulting polynomial with complex coefficients and
we lose crucial information.

We needed to prove a stronger version of the theorem:
\begin{coq}
Lemma mpolysym nu (p : {mpoly complexR[nu]}) :
  p \is a mpolyOver Cint -> p \is symmetric -> p != 0 ->
  {q | [&& (p == q \mPo [tuple 's_(nu, i.+1) | i < nu])
         , ((mweight q) <= msize p)
         & (q \is a mpolyOver Cint)]}.
\end{coq}
This proof is done by applying a morphism from polynomials with
complex-integer coefficients to polynomials with integer coefficients,
using the fundamental theorem in the type {\tt \{poly int\}} to obtain
a polynomial with integer coefficients, that can then be translated
into a polynomial with complex-integer coefficients.

The final transcendence statement has the following shape:
\begin{coq}
Theorem pi_transcendental : ~algebraicOver ratr PI
\end{coq}

\section{Conclusion and Future Work}

\subsection{Related work}
To our knowledge, the earliest formalized descriptions of
multivariate polynomials is proposed by Jackson \cite{Jackson:94}.  In a sense, this early
work already contains many of the ingredients that are found in our
description, with reliance on a free abstract monoid (while we rely on
a free group).  However, this early development contains very few
proofs.

Multivariate polynomials were also used by Théry \cite{Thery98} to prove that 
Buchberger's algorithm was  correct.  This work also relies on
an abstract description where each
polynomials is a sum of terms, where each term is the
multiplication of a coefficient and a monomial, the sums are
represented as ordered list of terms with non-zero coefficients.
Most of the operations, like addition or multiplication, take care of
maintaining the ordering and non-zero properties.  This made many of
the proofs rather complicated.  By
comparison, our development stays further away from implementation
details.  Another contribution fine-tuned for a specific algorithm is 
the work by Mahboubi on an efficient gcd algorithm
\cite{mahboubi:inria-00001270}.  Apparently, this description relies on a
recursive approach, where a \(n\)-variable polynomial is actually
encoded as 1-variable polynomial with 
\((n-1)\)-variable polynomials as coefficients.

More recently Mu{\~{n}}oz and  Narkawicz used multivariate polynomials in optimization problems \cite{MN13}.
The work by Haftmann et al.
\cite{haftmann14iw} is especially interesting because it enumerates
the various possible choices for the implementation of multivariate
polynomials and insists on the benefits of first implementing an type
of abstract polynomials.  It then shows how implementations can be
derived through refinements.  In a sense, the formal development that we
described here fulfills the requirements of an abstract type of
polynomials.  When considering
implementations, we also intend to rely on refinement
 described by Dénès, Cohen, and M{\"o}rtberg \cite{denes:hal-00734505,CohenDenesMortberg13}.

None of the work we have been able to find on formalized multivariate
polynomials includes any significant results about symmetric
polynomials and for this reason we believe that our work is the only one
to contain a proof of the fundamental theorem of symmetric
polynomials.

For formalized proofs of transcendence, the only prior work that we
are aware of is the formalized proof that \(e\) is transcendent
developed in Hol-light by Bingham \cite{JB11}.  The more complex proof
of transcendence for \(\pi\) was not formalized yet.  This is not
a surprise, since most known proofs use the fundamental
theorem of symmetric polynomials, for which we also provide the first
formalization.

\subsection{Future work}
This work started with an explicit aim of studying precisely the proof of
transcendence for \(\pi\).  We were able to isolate a
common lemma that applies both for the proof of transcendence
of \(e\) and \(\pi\).  As a continuation we are now considering
more general transcendence theorems, like Lindemann's theorem.
This theorem states results about linear and
algebraic independence of numbers and their exponentials.

We plan to extend the multivariate polynomials library to the case of
monoid algebras. While we expect minor modifications on the base
definitions and proofs ---notably on the ones for free groups that
are already abstracting over the set of generators--- this should
bring to the library the study of polynomials over of infinite number
of indeterminates, as well as the study of free modules. This
extension will allow us to fill gaps on some on-going formal
developments, notably in the proof of algorithms in the field of
algebraic combinatorics.

In the longer run, the infra-structure to combine the algebra library
Mathematical Components and the calculus library Coquelicot should be
improved.  Even though these libraries are developed with the same
system, we could probably re-use some of the tools that were developed
to communicate proofs from one theorem prover to the other.

\nocite{Coq}

\begingroup
\softraggedright

\endgroup

\end{document}

%% file: prelude.tex
\usepackage{xspace}

\newcommand{\Coq}[0]{Coq\xspace}
\newcommand{\ssr}[0]{SSReflect\xspace}

\usepackage[final]{listings}

\DeclareFontFamily{T1}{lmtt}{}
\DeclareFontShape{T1}{lmtt}{m}{n}{<-> ec-lmtl10}{}
\DeclareFontShape{T1}{lmtt}{m}{\itdefault}{<-> ec-lmtlo10}{}
\DeclareFontShape{T1}{lmtt}{\bfdefault}{n}{<-> ec-lmtk10}{}
\DeclareFontShape{T1}{lmtt}{\bfdefault}{\itdefault}{<-> ec-lmtko10}{}

\newcommand*{\ttfamilywithbold}{\fontfamily{lmtt}\selectfont}

\newcommand{\drarrow}{\raisebox{.07em}{$\scriptstyle\mathsf{-\mkern-3mu>\mkern-1mu}$}}
\newcommand{\dlarrow}{\raisebox{.07em}{$\scriptstyle\mathsf{<\mkern-3mu-\mkern-1mu}$}}

\iffinal
\fi
\def\ls{\lstinline}

\lstnewenvironment{coq}{%
\lstset{basicstyle=\ttfamilywithbold\small}
\lstset{belowskip=1em,aboveskip=1em}}{}